\documentclass{PoS}
\usepackage{graphicx}
\usepackage{amsmath,amssymb,amsfonts}
\usepackage{sidecap}

\title{Quantum insights on Primordial Black Holes \\ as Dark Matter}

\ShortTitle{Quantum insights on Primordial Black Holes as Dark Matter}

\author{\speaker{Francesca Vidotto}\thanks{
The work presented is the fruit of collaborations, in particular with Aurelien Barrau, Alvise Raccanelli, and Carlo Rovelli.
The research of FV at UPV/EHU is supported by the grant IT956-16 of the Basque Government and by the grant FIS2017-85076-P (MINECO/AEI/FEDER, UE).
}\\
        University of the Basque Country UPV/EHU, Departamento de F\'isica Te\'orica, \\ Barrio Sarriena s/n, 48940 Leioa, Biscay, Spain\\
        E-mail: \email{francesca.vidotto@ehu.es}}

\abstract{A recent understanding on how quantum effects may affect black-hole evolution opens new scenarios for dark matter, in connection with the presence of black holes in the very early universe. Quantum fluctuations of the geometry allow for black holes to decay into white holes via a tunnelling. This process yields to an explosion and possibly to a long remnant phase, that cures the information paradox. Primordial black holes undergoing this evolution constitute a peculiar kind of decaying dark matter, whose lifetime depends on their mass $M$ and can be as short as $M^2$. As smaller black holes explode earlier, the resulting signal have a peculiar fluence-distance relation. I discuss the different emission channels that can be expected from the explosion (sub-millimetre, radio, TeV) and their detection challenges. In particular, one of these channels produces an observed wavelength that scales with the redshift following a unique flattened wavelength-distance function, leaving a signature also in the resulting diffuse emission. I conclude presenting the first insights on the cosmological constraints, concerning both the explosive phase and the subsequent remnant phase.
}

\FullConference{2nd World Summit on Exploring the Dark Side of the Universe,\\
		25-29 June 2018\\ Guadeloupe Islands}

\begin{document}

\maketitle

\section{Primordial Black Holes} \label{intro}
Dark matter is maybe the most puzzling open problem, brought to physicists by mounting observational evidences in astronomy. While for long the idea of new particles dominated the theoretical landscape, on the experimental side, where the search effort spanned from accelerator experiments to space telescopes, no evidence for dark matter particles has manifested. This situation encourages us to explore solutions that involve gravitational physics rather than the presence of new particles. 

Since the detection by the LIGO interferometer of black holes in an unexpected mass range, Primordial Black Holes (PBHs) have experienced a renovated interest, that seems so far steadily increasing. 
According to a variety of scenarios, PBHs are produced in the radiation-dominated era, and they are a possible candidate for dark matter (DM). 
PBHs are constituted by matter that is not subject to big-bang nucleosynthesis constraints on baryons, therefore they  behave to all effect as non-baryonic dark matter (DM). The idea that PBHs could contribute to DM dates back to the Seventies \cite{Chapline:1975cr}  
 and since then observational consequences and constraints have being considered for different PBH mass ranges exploiting different observables \cite{Carr:2016hva,Chen:2016pud,Green:2016xgy}.
The search for PBHs and for constraints on PBHs has mostly considered an almost monochromatic mass spectrum, and the presence of Hawking evaporation for PBHs of small mass.

A monochromatic mass spectrum has been challenged by different authors as restrictive and unrealistic \cite{Carr:2017jsz}. 
 An extended mass function is compatible with different PBH formation mechanics, from critical collapse to cosmic strings.  
 An extended mass function, encompassing very massive one (like the ones detected by LIGO and above) and sub-solar ones as well, makes more viable the possibility to saturate DM with PBHs. 
 
Hawking evaporation is a phenomenon that becomes relevant on a time scale that depends on the mass of the black hole.  In Planck units, the evaporation time scale is $M_{BH}^3$. This implies that within the age of the universe only PBHs with mass smaller than $10^{12}$ Kg could have evaporated, and possibly produced very high-energy cosmic rays~\cite{Barrau:1999sk}: as cosmic rays of such energies are rare, constraints are derived on the very-small-mass end of the PBH mass spectrum.

Hawking evaporation, however, is a phenomenon predicted in the context of quantum field theory on a fixed curved background. This is a theory with a regime of validity that is likely to break down when approximately half of the mass of the hole has evaporated, as indicated for instance by the `firewall' no-go theorem \cite{Almheiri:2012rt}. The geometry around a black hole can indeed undergo quantum fluctuations on a time scale shorter than $M_{BH}^3$, when the effects of the Hawking evaporation have not not yet significantly  modified the size of the hole. As any classical system, the hole has a characteristic timescale after which the the departure from the classical evolution become important as quantum effects manifest. This time can be much shorter than the Hawking evaporation time $M_{\rm BH}^3$, as short as $M_{\rm BH}^2$: this is the minimal time after which quantum fluctuations of the metric can appear in the region outside the hole horizon. 
In fact, the presence of a small curvature proportional to $M_{\rm BH}^{-2}$ near the hole horizon allows to define the `quantum break-time' as the inverse of this quantity, i.e. $M_{\rm BH}^2$ \cite{Haggard:2014fv,Haggard:2016ibp}. As soon as these quantum fluctuations of the geometry take place, the dynamics of the horizon can undergo dramatic behaviours, to the point of the very disappearance of the horizon,  possibly by an explosion.

\section{Quantum Gravity}

The classical curvature singularity at the center of BHs is cured in the quantum theory by non-perturbative effects. 
In Loop Quantum Gravity singularity resolution has been studied extensively
\cite{Rovelli:2013osa,Ashtekar:2005qt,Corichi:2015xia}.
Quantum discreteness plays a fundamental role: this is not just a space discreteness, but a \emph{spacetime} discreteness, that modifies both the kinematics and the dynamics of the theory. 
The quantum equations of motions can be translated into effective equations, that provide a compelling intuition of how quantum effects concretely affects black hole evolution: an effective repulsive force appears when the energy density approaches Planckian values, preventing any further collapse. Instead of the singularity, the repulsive force triggers a new expanding phase. 

A non-singular black hole forms as usual in the collapse of some matter and it is associated with the accompanied by the formation of a horizon. This scenario is rather conservative with respect to other model of non-singular black holes, because the collapse do produces a horizon. On the other hand, this is not an event horizon but a dynamical horizon with a finite lifetime, hence it can be fully characterised as a \emph{trapping surface}.
The collapse continues beyond this surface and reaches a maximal contraction: this phase of the collapse is referred in the literature with the suggestive name of {\it Planck Star} \cite{Rovelli:2014cta}. At this point the appearance of an effective quantum pressure 
starts the expansion and the consequent creation of an \emph{anti-trapping surface}, i.e. a white hole.
The passage from a contacting solution to an expanding one is called a \emph{bounce}. Notice that the bounce connects two classical solution of the Einstein equations \cite{Haggard:2014fv} via a quantum region. This is the definition of a quantum tunnelling, a characteristic non-perturbative quantum phenomenon. In analogy with nuclear physics, we can then refer to the black-to-white tunnelling as a decay, and discuss the hole lifetime, that characterises the time-scale of the process.

Notice that two time-scales should be distinguished. One is associated with the proper time of the collapsing matter that forms the BH: seen from the center of the hole, the matter collapses trough the dynamical horizon and then expands trough the horizon (now a white-hole horizon). Consider the case of BHs formed in the early universe when this matter was basically constituted by photons: the bounce will then happen at the speed of light, and last roughly the time light takes to cross the region inside the horizon (this is of the order of micoseconds for a BH of solar mass). A different time scale is associated to an observer situated outside the horizon: this is extremely long due to the huge (but finite) redshift. This is the reason why we do see astrophysical black holes in the sky and they behave for us as stable objects!

On the other hand, primordial black holes are small and old enough to have an observable decay for observers situated outside of the horizon as we are. From the discussion in Sec.\ref{intro} we conclude that the black-hole lifetime should last, in natural units, for a time between  $M_{\rm BH}^2$ and $M_{\rm BH}^3$. Standard decay physics suggests that the shortest value $M_{\rm BH}^2$ is the one most probable.
As the decay is a fully non-perturbative process, a non-perturbative quantum theory of gravity is necessary to compute the hole lifetime. This computation provides a possible terrain to discriminate between different quantum-gravity approaches. While instabilities leading to a black-hole decay are present in different approaches \cite{Gregory:1993vy,Kol:2004pn}, so far only Loop Quantum Gravity provides sufficiently-defined tools to compute explicitly this lifetime \cite{Christodoulou:2016ve}.

For the sake of the phenomenology, it is possible to study the BH decay considering a parametrisation that spans the whole window of allowed lifetimes  between $M_{\rm BH}^2$ and $M_{\rm BH}^3$. Then we can write the 
BH lifetime $\tau$ with a parameter $\kappa$ as:
\begin{equation} \label{eq:alpha}
\tau = 4\kappa\left(\frac{M_{\rm BH}}{m_{\rm Pl}}\right)^{2} t_{\rm Pl} \, ,
\end{equation}
where $ t_{\rm pl } $ and $ m_{\rm pl } $ are the Planck time and the Planck mass.

\section{Phenomenology}

This scenario leads to a number of consequences with observational significance, listed below.  

\vskip1mm \hspace{-5mm}
$\bullet$\ \     
If PBHs form a component of DM, this would be DM decaying with a mass-dependent decay time --a peculiar feature as other decaying DM candidates have a fixed decay time. As the decaying time $\tau$ is shorter than the evaporation time, the effects of such a quantum-gravity phenomenon dominate over Hawking evaporation; this leads to {\em superseding the constraint on PBHs as DM obtained from the Hawking evaporation}.

\vskip1mm \hspace{-5mm}
$\bullet$\ \      
The scenario leads to a threshold in the minimal mass of PBHs present today. In the case of Hawking evaporation, a minimal mass was set by the fact that PBHs evaporating today have a mass equal to $10^{12}$ Kg: those of smaller mass have had already evaporated. In the case of a quantum-gravitational decay, the mass of those exploding today can be as high as $10^{23}$ Kg for the shorter lifetime $M_{\rm BH}^2$. Remarkably, this {\em relaxes the constraints from microlensing of object smaller than such mass }\cite{griest2013new}.

\vskip1mm \hspace{-5mm}
$\bullet$\ \      
PBH decay has the peculiar property of lowering the DM energy-density content of the Universe as the decay converts effectively DM into radiation, modifying the equation of state of the Universe though ages. This {\em affects the galaxy number count in large-scale galaxy surveys, in particular measuring the galaxy clustering, galaxy lensing and Redshift-Space Distortions (RSD)}    \cite{Raccanelli:2017one}. The LSS surveys by SKA will provide key data in this respect by detecting individual galaxies in the radio continuum~\cite{Jarvis:2015ska}. As we are interested in how these data may evolve with redshift while the SKA survey do not measure directly the redshift, we can apply different techniques such as the division of the galaxy catalog into tomographic redshift bins \cite{Menard:2013aaa,Kovetz:2016hgp}.
~
These measurements acquire particular interest if PBHs will be detected using different observables: then the quantum-gravity lifetime-mass scaling would be constrained from this type of analyses, opening the possibility to measure a quantum-gravity phenomenon from late-universe data.

\vskip1mm \hspace{-5mm}
$\bullet$\ \     
Effects of the quantum-gravitational BH decay could be identified {\em in the CMB from the energy injection by the PBH explosion}; to understand further the cosmological implication of the BH decay future research should aim at investigating these effects and developing a constraint analysis combining different observables, taking into account the possibility of an extended PBH mass function.

\vskip1mm \hspace{-5mm}
$\bullet$\ \     
{\em Astrophysical signals produced in the explosive event associated to the BH decay could be  detectable directly.}   Lacking of a detailed astrophysical model for the BH explosion, emission channels can be studied on the basis of heuristic arguments. Two channels can be expected: (\ref{HEC}) a signal determined by the temperature of the photons emitted (high-energy channel), (\ref{Rees}) that in presence of magnetic fields 
can produce a signal in the radio (Rees' mechanism), and (\ref{LEC}) a signal determined by the size of the hole exploding (low-energy channel). and (iv) the emission in gravitational waves. The main parameter that determines the specific features of all these signal is the BH mass. We will now examine these channels separately. 

A further question is the possible gravitational-wave production via the black-to-white bounce:  we omit this here, even if in the era of multimessanger astronomy this question is particularly interesting and deserve further investigations.

\subsection{High Energy Channel}\label{HEC}
The \emph{high energy} channel is constituted by the matter, primarily photons, re-emitted with the same temperature it had at the time of the collapse that formed the PBH. In the original collapse model \cite{Carr:1974pbh}, during the reheating over-dense region can collapse forming PBHs with different sizes determined by the Hubble horizon. As the temperature at reheating is of the order of the $TeV$, the energy of the photons in the high energy channel is expected in this range. New instruments should became available in the forthcoming years to observe in this energy window \cite{Rieger:2013awa}. Observing signals in this  channel is limited by a significative observational horizon: at such a high energy, cosmic rays interact with the CMB, allowing the signal to travel to us only when the source is located within our galactic neighbourhood. See \cite{Barrau:2015acl} for the analysis of the signal detectability.
\subsubsection*{Rees' Mechanism and Fast Radio Bursts}\label{Rees}
The study of high-energy cosmic emissions from PBH explosions was initiated in the Seventies and the problem detecting sources at a cosmological distance was immediately realised. Martin Rees proposed a mechanism \cite{Rees:1977nat} that allowed to circumvent the problem and make most of the PBH explosions in the visible universe virtually accessible. 
He noticed that, in presence of a ionised interstellar medium, exploding PBHs can produce a secondary radio pulse of the order of $\sim1GHz$.
The mechanism relies on the production during the explosion of a shell of relativistic charged particles: the shell behaves as a superconductor that expels the interstellar magnetic field from a spherical volume centred in the original BH sites.  The structure of the resulting secondary radio signal was investigated in details in \cite{Blandford:1977}.

Interestingly, the scenario discussed in this paper gives a crucial difference with respect to the old models of PBH. 
In fact, PBH explosions at the end of the Hawking evaporation were expected to emit a high energy burst, where the total mass of the object converted into the burst was just Planckian. This was the case for all exploding PBHs, independently from their original mass. The bursts were therefore expected to have all the same fluence, and the fluence to be very small - an aspect that made more difficult the detection.
~%
On the other hand, our scenario presents a different characteristic feature. The lifetime of the blackhole is a function of its mass and the explosion can be expected before the Page time, when the black holes has conserved a mass of the same order of magnitude of when it formed. 
PBHs observed exploding far away are those less massive, and they producing a primary emission of lower fluence than those at a closer distance. The Rees' mechanism is very efficient in transforming 
the primary emission into the secondary radio emission, almost preserving the original fluence, therefore the fluence of the radio burst carries the information about the original mass of the PBH source. Therefore 
a characteristic fluence-distance relation for the radio signal can single out PBH decays with respect to other astrophysical  sources measured at cosmological distance.

Fast Radio Bursts (FRB) have been discovered a decade ago and their origin is today still an open question. BH exploding via a black-to-white transition constitutes a natural candidate as FRB source. FRBs presents a number of singular features that can be easily matched by the model presented here. They are extremely energetic impulsive events, at cosmological distance. Their magnitude of their enormous fluence can be explained by the conversion of the whole PBH mass into radiation \cite{Barrau:2014frb}. Recent studies seems to indicate that FRBs follow a non-standard fluence-distance relation \cite{James:2018frb,Shannon:2018frb}, a fact that future radio surveys with higher sensitivity can help to clarify \cite{SKA:2018}. Finally, the FRB signal is polarised, hinting to the presence of a magnetic field at the source \cite{Petroff2014,Akahori2016,Lu2018}.  

\begin{figure}[b]
\label{fig:flat}
\centering
\includegraphics[height=55mm]{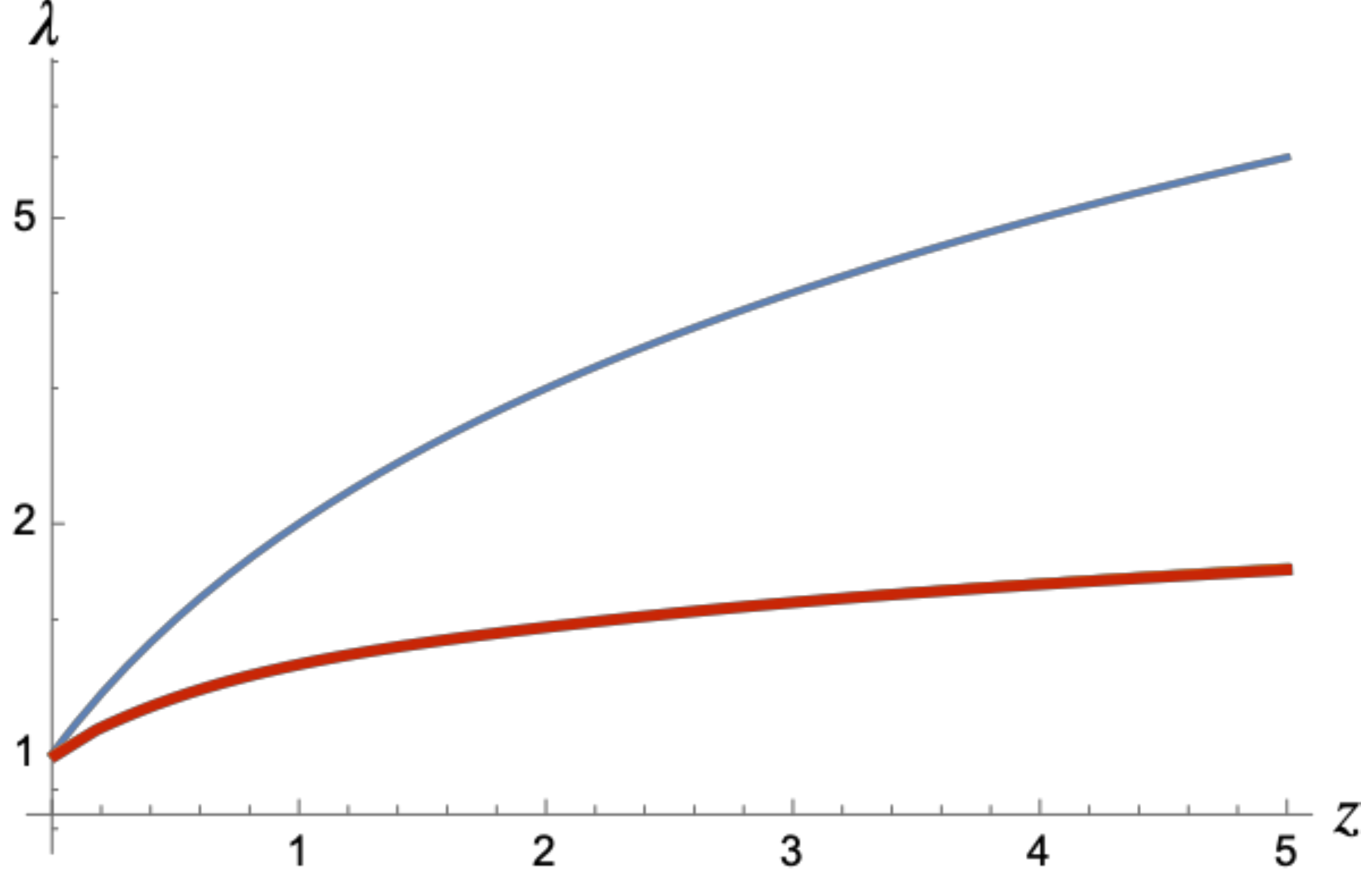}
\caption{In blue, the dependence of the observed wavelength from the redshift $z$ for a standard astrophysical object. In red, the expected wavelength (unspecified units) of the signal from black hole explosions as a function of the redshift $z$. The curve flattens at large distance: the shorter wavelength from smaller black holes exploding earlier get compensated by the redshift. }
\end{figure}

\subsection{Low energy channel}\label{LEC}

The \emph{low energy} channel is more promising in terms of detectability. This emission channel should be present if the explosion excites a mode correspondent to the BH size, given roughly by the Schwarzschild radius.  This depends on the actual value of the lifetime. Using \eqref{eq:alpha} we can indeed compute the mass of black holes whose lifetime is the Hubble time, yielding an explosion today. For a lifetime in the lower end of the viable window, the Schwarzschild radius is of the order of  $R_{BH}\sim.2mm$ and the observational dept allows to detect events in all the visible universe. 

For the highest values of $\kappa$, the signal is emitted with energy in the $GeV$, but a study of the photon emission with the code PHYTIA indicates that the photons with higher density, therefore more likely to be detected, are those in the $MeV$ \cite{Barrau:2014phe}. Some Short Gamma Ray Bursts \cite{Nakar:2007yr}, whose source is still unclear, are possible candidates in this range.

For the lowest values of $\kappa$, the peak of the signal in the millimetres. The corresponding frequencies are beyond the sensitivity of most radio telescopes, such as SKA-mid. On the other hand, in a decay we expect a probability distribution of the event to happen, therefore there could be some event happening within SKA-mid frequency range. Furthermore, the details of the decay mechanism are not fully established, leaving open the possibility of the presence of some factors that could shift the wavelength possibly of 1-2 orders of magnitude. This makes worthy to explore for transients the highest frequencies accessible to SKA-mid. In particular, it has been suggested that the quantum-gravitational Bh explosion may be linked to Fast Radio Bursts  \cite{Barrau:2014bv}. A number of hints seems to support this conjecture, such the rapid impulse, a mostly extra-galactic origin, the enormous energy flux as the BH converts efficiently its mass into radiation.

\subsubsection*{Signature of PBH decay} 
PBH exploding in a quantum decay present a unique feature that makes them distinguishable from other astrophysical sources. The time at which the decay happens is a function of the BH mass. Therefore, 
smaller PBH explode earlier, i.e. at a distance from us. The signal they produce also depends on their mass: the smaller black holes, the shorter wavelength in the high and in the low energy channels. Signal from distant sources get redshifted, partially compensating the shorter wavelength due to a smaller mass. 

Generic astrophysical objects have an observed wavelength that scale linearly with the distance. On the other hand, Hawking evaporation ends in the standard theory when BHs, of any initial mass, reach the Planck size and produce a signal with such wavelength. The phenomenon described above gives a completely different effect: the modified relation between wavelength and redshift \cite{Barrau:2014frb} represented by the flattened curve (Fig.\ref{fig:flat}). Observation of burst whose source has a known distance will allow to see whether data fit such a curve. If so, it would be a signature of the quantum-gravitational nature of the process at the origin.

An exact localisation of burst sources could be proven difficult. This involve the dispersion measure of the received signal, and possibly the source being hosted in a known astronomical object, for instance a galaxy. A further technique is available, and it is peculiar of decaying PBHs. BH exploding far away are less massive, with a lower flux of energy: 
 the flux of energy of the measured burst lessen  with a distance.
\begin{figure}[b]
    \centering
        \vskip1mm 
        \includegraphics[width=.47\textwidth]{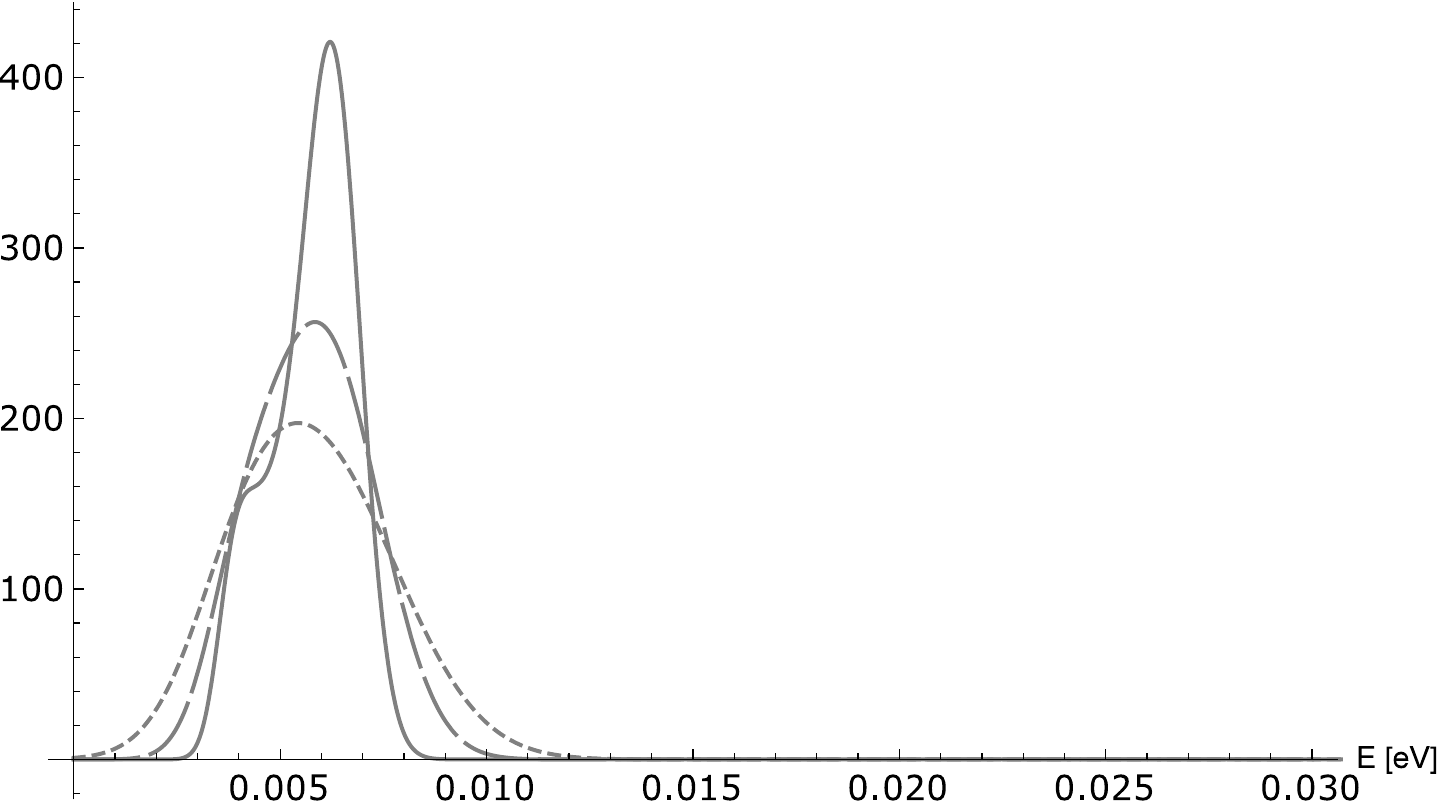}
        ~~~~
        \includegraphics[width=.47\textwidth]{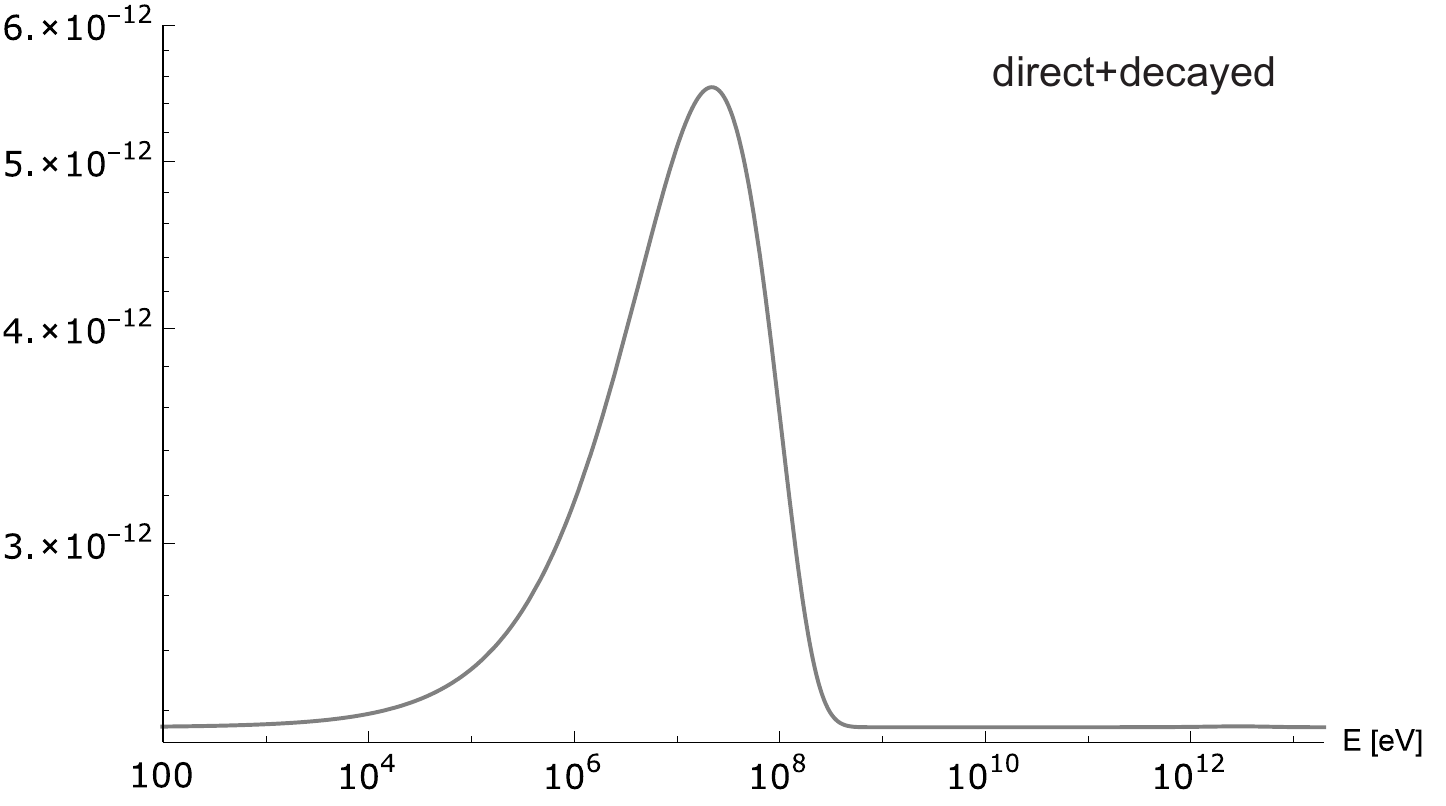}
\caption{The diffuse emission of the low energy channel, and below the diffuse emission of the high energy channel. We show here the ones for the shortest BH lifetime, i.e. the lowest $\kappa$ parameter. The spectrum was studied for the full range of $\kappa$ in \cite{Barrau:2015acl}. The units in the ordinate axis are not specified: the normalisation of the spectrum depends on the percentage of PBHs as DM.}
\label{fig:diffuse}
\end{figure}
\subsubsection*{Intensity mapping} 
The peculiar wavelength-redshift relation can be proven also without the knowledge of the distance of the source, by studying the radiation integrated over the detectable past explosions. The instruments performing intensity mapping campains can therefore provide the desired data in this respect. The study the integrated emission from PBH decay has been started in \cite{Barrau:2015acl} for a large range of the $\kappa$ parameter in the PBH lifetime. The signal has been obtained using the PYTHIA code \cite{Sjostrand:2014zea}, that for a given initial energy computes the particle production for the process.
The resulting radiation is not thermal, but it carries a distortion due to the characteristic redshift-wavelength relation of Fig.\ref{fig:flat}. This appears for both the high energy and the low energy channel, as can be seen in Fig.\ref{fig:diffuse} for the case of the shortest lifetime. 
In principle the result depends on the PBH mass spectrum, but different hypothesis for the mass spectrum have been tested showing that the result have only a very weak dependence \cite{Barrau:2015acl}.

\section{PBH remnants and dark matter}
Quantum gravity provides a concrete mechanism to describe the final stage of the life of the black hole, that does not finish at the end of the evaporation nor with the explosion described above. In fact, the black-to-white transition allows for a new remnant phase, 
characterised by
with a residual small Planck mass $\sim10^{-9}\ Kg$.

Two observations support the idea that these remnants are stable on timescales much longer than the age of the universe.
The first observation is based on classical General Relativity. When a black hole forms, it has a large internal volume. Hawking evaporation or macroscopic black-to-white decays are dramatic events that changes the size of the horizon surface; on the other hand, the ``reabsorption'' of the large internal volume can only happens on a longer timescale $\sim M_{BH}^4$, where $M_{BH}$ is the original mass of the black hole rather than the final remnant mass \cite{Bianchi:2018rem}.~
The second observation uses a basic prediction of Loop Quantum Gravity. Spacetime quantization yields a discrete spectrum for geometrical quantities. In particular, at the Planck scale an area measurement reveal the presence of a minimal non-zero eigenvalue.~

The viability of these remnants as a light dark matter component has been proposed in 
\cite{Rovelli:2018awo}. Two possible scenario are currently investigated: one in which PBHs are formed at the reheating time \cite{Rovelli:2018dmr}, and one where PBHs are formed in a bouncing cosmology \cite{Rovelli:2018ple}. The latter scenario is particularly intriguing because BH can act as a dust component that dominates the contracting phase of a bouncing universe, giving an almost scale-invariant power spectrum even in absence of inflation \cite{Wands1999,Brandenberger:2012zb}.

\bibliographystyle{utcaps}
\bibliography{/Users/francesca/Library/Texmf/tex/bib/library}

%

\end{document}